# Graph Structure Learning for Tumor Microenvironment with Cell Type Annotation from non-spatial scRNA-seq data

Yu-An Huang[1,2,#,*], Yue-Chao Li[1,#], Hai-Ru You[3], Jie Pan[4], Xiyue Cao[1], Xinyuan Li[1], Zhi-An Huang[5,*], Zhu-Hong You[1,*]

[1] School of Computer Science, Northwestern Polytechnical University, Shaanxi 710129, China

[2] Research & Development Institute of Northwestern Polytechnical University in Shenzhen, Shenzhen, 518063, China

[3] School of Engineering, The University of Sydney, Darlington, New South Wales 2037, Australia

[4] College of Life Sciences, Northwest University, Xi'an, 710069, China

[5] Research Office, City University of Hong Kong (Dongguan), Dongguan 523000, China

#equal contribution

*corresponding authors

## Abstract

The exploration of cellular heterogeneity within the tumor microenvironment (TME) via single-cell RNA sequencing (scRNA-seq) is essential for understanding cancer progression and response to therapy. Current scRNA-seq approaches, however, lack spatial context and rely on incomplete datasets of ligand-receptor interactions (LRIs), limiting accurate cell type annotation and cell-cell communication (CCC) inference. This study addresses these challenges using a novel graph neural network (GNN) model that enhances cell type prediction and cell interaction analysis. Our study utilized a dataset consisting of 49,020 cells from 19 patients across three cancer types: Leukemia, Breast Invasive Carcinoma, and Colorectal Cancer. The proposed scGSL model demonstrated robust performance, achieving an average accuracy of 84.83%, precision of 86.23%, recall of 81.51%, and an F1 score of 80.92% across all datasets.

These metrics represent a significant enhancement over existing methods, which typically exhibit lower performance metrics. Additionally, by reviewing existing literature on gene interactions within the TME, the scGSL model proves to robustly identify biologically meaningful gene interactions in an unsupervised manner, validated by significant expression differences in key gene pairs across various cancers. The source code and data used in this paper can be found in https://github.com/LiYuechao1998/scGSL.

**Keywords:** Single-cell RNA sequencing, Tumor Microenvironment, Cell Type Annotation, Cell-Cell Communication, Ligand-Receptor Interactions

## 1. Introduction

Single-cell RNA-sequencing (scRNA-seq) technology has revolutionized our understanding of the cellular heterogeneity within the TME, a complex ecosystem integral to cancer progression and therapeutic response. The TME consists of a diverse array of cellular components including immune cells, endothelial cells, and stromal cells, alongside molecular components such as signaling molecules and extracellular matrix proteins. These elements do not exist in isolation. Instead, they engage in extensive cross talk that significantly influences tumor behavior and patient outcomes. Through scRNA-seq, the intricate cellular interactions and functional heterogeneity that traditional bulk sequencing methods often obscure are now illuminated, providing insights into the unique transcriptional profiles of individual cells. This intricate detail enables the identification of specific subpopulations of immune cells that may be uniquely associated with either tumorigenesis or tumor suppression. It also delineates the role of stromal cells like fibroblasts in supporting tumor growth through the secretion of matrix metalloproteinases (MMPs) and other enzymes that facilitate tumor invasion and metastasis. By offering a clearer view of the cellular landscape within the TME, scRNA-seq facilitates a deeper understanding of how these diverse cell types contribute to cancer dynamics and resistance, potentially leading to more effective therapeutic strategies.

However, a significant challenge of scRNA-seq is its inability to retain spatial context of the cells, necessitating subsequent computational analysis for cell type annotation and the inference of intercellular communication. These analyses are crucial for constructing a comprehensive map of the cellular interactions within the TME, which can reveal potential targets for therapeutic intervention. By understanding how different cell types within the TME communicate and contribute to cancer progression, scRNA-seq enhances the understanding of cancer biology and opens up new avenues for targeted therapy.

While scRNA-seq captures gene expression at the individual cell level, the methods for analyzing CCC predominantly focus on interactions between cell types rather than individual cells. These methods rely fundamentally on the introduction of LRIs to score signaling communication between different cell types. LRIs involve a ligand, typically a protein or small molecule secreted by one cell, binding to a specific receptor on another cell. This binding initiates a cascade of signals within the receiving cell, thereby facilitating communication between cells. LRIs are crucial for explaining how cells influence each other in biological contexts, forming the basis of intercellular communication.

As summarized by Peng et al. [1], existing computational approaches for inferring CCC are categorized into three main types: (i) network-based scoring methods, including CCCExplorer [2] and NicheNet [3]; (ii) machine learning prediction techniques, such as PyMINEr [4],

SoptSC [5], and SingleCellSignalR [6]; and (iii) spatial information-based inference tools like CellTalker [7], SpaOTsc [8], and histoCAT [9]. The computational pipeline for these methods typically begins with cell type annotation or cell clustering, followed by the organization of original LRI data. Gene expression values of ligands and receptors are used to compute interaction scores for each pair mediating between two cell types, and these scores are aggregated to assess the overall state of communication between them [10]. Each method has its strengths and weaknesses when compared: network-based scoring and machine learning approaches heavily rely on the curation of LRIs and their downstream pathways, whereas spatial information-based inference is limited to spatial datasets and cannot be applied to non-spatial transcriptomics data. CCC inference is increasingly viewed as a data-driven analysis task that varies with different techniques and relies on diverse scoring approaches to determine cell type specificity, yet the absence of a "gold standard" dataset makes it challenging to evaluate and standardize these methods.

However, existing CCC inference methods universally confront several significant issues. The acquisition of LRIs relies heavily on biological experiments, leading to an incomplete dataset where some interactions are missing, such as ligands like IL17D that lack known receptors and receptors like RELT, NGFR, and TROY from the TNF receptor family that lack known ligands. This incompleteness can introduce biases, particularly affecting rare cell types. Moreover, LRIs are not tumor-specific, implying that in TME scenarios, all known LRIs are considered equally critical across all cell types, potentially leading to an overemphasis on certain interactions. Most CCC inference techniques focus on proteinaceous ligands and receptors, neglecting other important signaling cofactors such as stimulatory and inhibitory membrane-bound co-receptors, along with soluble agonists and antagonists. Beyond the shortcomings related to LRIs, another challenge lies in the dependency of CCC inference methods on cell-type identification, heavily relying on tools like Seurat [11], SingleR [12], Cellassign [13], and scCATCH [14]. However, these methods have an accuracy rate of about 80% and still require significant improvement.

Recent advances in cell-type identification have seen methods that construct cell-cell graphs and utilize GNNs to tackle the task of node label prediction for cell type annotation. For example, scGCN [15] constructs its graph by calculating similarities between cell expression profiles and applying sparsification. scGNN incorporates interaction probability scores derived from the CellChat [16] method and LRIs to assign weights to the edges of its graph. scPML [17] calculates edge weights using AUCell [18] scoring of gene signaling pathways from databases like KEGG [19]. Despite the differing methodologies in constructing graphs, these approaches have confirmed that establishing links between cells is beneficial for predicting cell types.

In conclusion, computational methods for CCC inference are heavily reliant on the predictive performance of cell type annotation. Concurrently, methods for cell type annotation are increasingly incorporating considerations of inter-cell interactions within CCC. Yet, the graphs constructed by graph-based cell type annotation methods, while serving as inputs for GNNs, have rarely been explored for their interpretability. Inspired by this observation, there is a reciprocal benefit to be realized between cell-cell interaction and cell type annotation. To leverage this potential, we propose a new computational model based on GNNs. Utilizing unlabeled scRNA-seq data from tumor tissues, our model addresses two pivotal computational tasks essential for analyzing the TME: firstly, the annotation of cell types, and secondly, the construction of a graph that measures potential interactions among cells within the TME.

Unlike existing GNN-based computational methods for single-cell type prediction, which construct the cell-cell graph through filtering. Our hypothesis posits that enhancing the graph representation of individual cells could significantly improve both the predictive accuracy and interpretability of cell type identification. By developing a more sophisticated neural network model that can learn these complex interactions, we aim to not only boost the predictive performance but also gain insights into the cellular mechanisms at play. Consequently, this allows for the exploration of the TME at a more detailed, single-cell level of interaction, rather than the broader cell type interactions typically analyzed in traditional studies with non-spatial scRNA-seq data.

## 2. Methodology

### 2.1 scRNA-seq dataset

In this study, we utilized a comprehensive scRNA-seq dataset to predict cell types across various diseases. The dataset encompasses a total of 49,024 cells derived from 19 individuals, averaging approximately 2,580 cells per individual. This rich dataset includes cells from patients suffering from three distinct diseases: Leukemia, Breast Invasive Carcinoma, and Colorectal Cancer, which are further subdivided into specific datasets as described below and summarized in Fig. 1, a bar graph illustrating the distribution of cell counts and cell type proportions among the patients. Specifically, this dataset collection encompasses leukemia samples from GSE142213 (including AYL050 and OX1164) [20], GSE132509 (including PBMMC_1, PBMMC_2, PBMMC_3) [21], and GSE154109 (including P6 and P7); breast invasive carcinoma samples from GSE143423 (including tnbc1 and tnbc6k) [22], E-MTAB-8107 (including sc5rJUQ024, sc5rJUQ026, sc5rJUQ033, sc5rJUQ050, sc5rJUQ060) [23], and GSE150660 (including GSM4555888, GSM4555891) [24]; and colorectal cancer samples from E-MTAB-8107 (including scrEXT009, scrEXT010, scrEXT020).

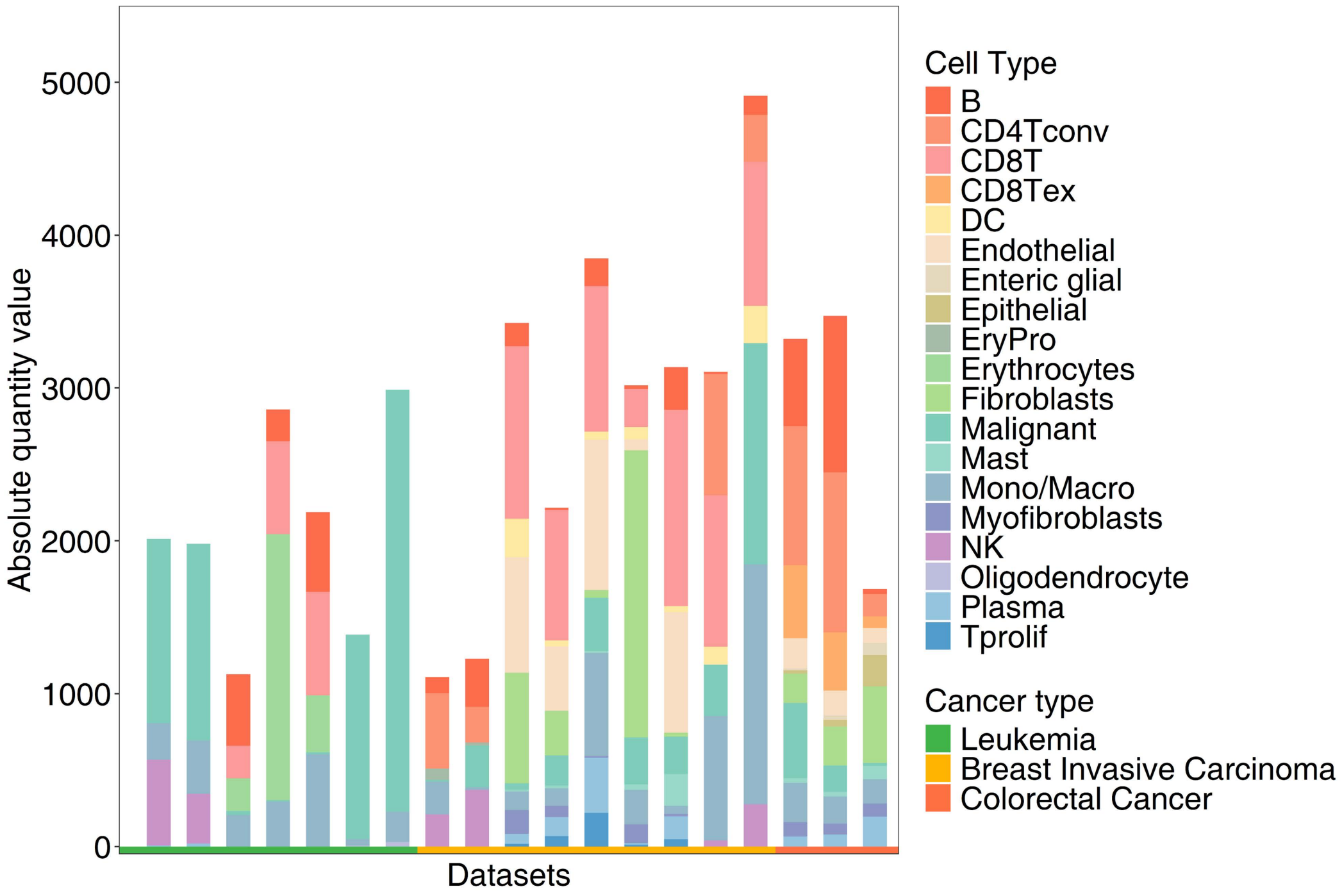

**Fig. 1** Statistical Summary of scRNA-seq Dataset.

## 2.2 Overview of scGSL model

In this work, we present scGSL, a new GNN-based model geared for analyzing non-spatial transcriptomic scRNA-seq data within cancer research contexts. scGSL utilizes labeled cell type data from scRNA-seq of a particular cancer to annotate other datasets of the same cancer type and compute the cell-cell interaction graph. Notably, scGSL integrates two advanced computational strategies: GSL and graph domain adaptation [25]. GSL is employed to autonomously derive a representative cell-cell graph based on the intrinsic labels available from the data, thereby enhancing the graph's capacity to accurately reflect biological interactions. Meanwhile, graph domain adaptation functions within the non-Euclidean space of graphs to adapt domains, thereby enabling classifiers trained on a reference dataset to more effectively predict cell types in a query dataset of the same cancer type.

The overall workflow of scGSL is depicted in Fig. 2 and consists of three main stages: (1) The initial stage involves transforming scRNA-seq matrix data from both reference and query sets into cell-cell graphs. For the reference set, the graph construction starts with KNN and is further refined using GSL, employing cell type labels for supervision, resulting in a sophisticated cell-cell graph where individual cells are represented as nodes and relationships defined through GSL; for the query set, the graph is solely constructed using KNN. Detailed methodology is discussed in Section 2.3. (2) The second stage builds upon this graph structure, using cell type labels from the reference dataset to perform supervised training of a GNN classifier. This process is enhanced by a graph domain adaptation framework known as pairwise alignment which tailors the classifier for efficient prediction on the query dataset of the same cancer type. (3) The final stage focuses on utilizing the predicted cell types and learned embeddings from the query dataset as new node features and labels, respectively. These are subsequently re-input into the same GSL module used in the first stage to learn the cell-cell graph of the query dataset. This aims to reveal the complex cellular interactions within the TME.

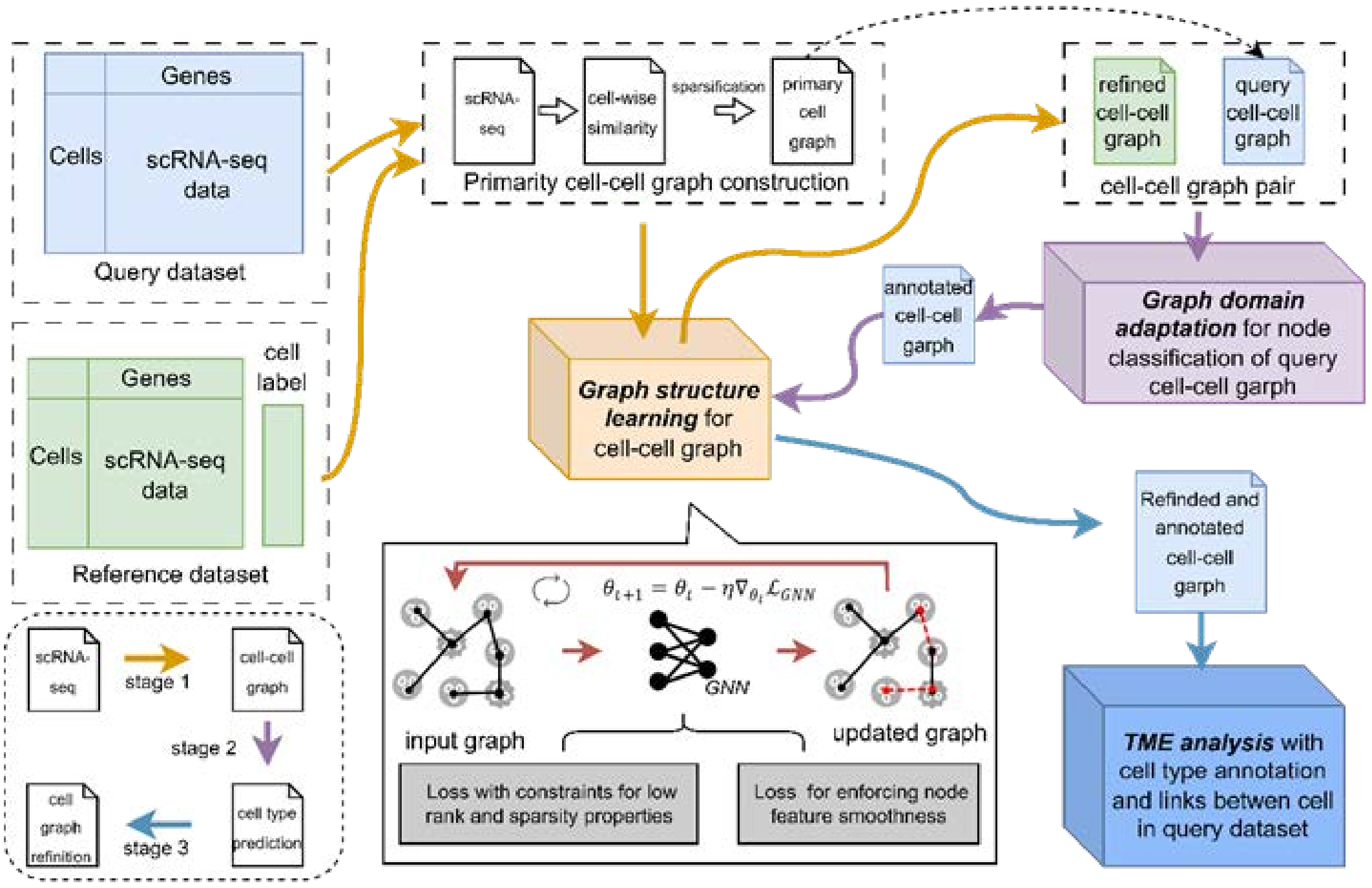

**Fig. 2** Flowchat of scGCN model.

## 2.3 Graph structure learning for cell-cell graph

Upon acquisition of scRNA-seq data, in this work, rigorous quality control was implemented by removing cells with gene expressions below 200 and genes expressed in fewer than three cells, ensuring reliability in subsequent analyses. Common genes were then identified between source and target datasets, and the top 3000 highly variable genes (HVGs) were selected to highlight biological variability. To address batch effects, normalization procedures were applied to align features across datasets. The KNN algorithm, setting neighbors at 0.2% of total cell count, was used to construct graphical representations for both datasets.

In constructing cell-cell interaction graphs from non-spatial transcriptomic scRNA-seq data, challenges such as drop-out events, amplification bias, and insufficient sequencing depth inherently distort the representation of genuine cellular interactions. These issues introduce noise and perturbations that significantly complicate the interpretation of calculated correlations in gene expression profiles, as well as scores derived from signaling pathways or LRIs. We propose that these adverse effects can be mitigated by incorporating cell type information and by preserving two intrinsic characteristics typically observed in real-world graph data: feature smoothness in gene expression profiles and the low-rank and sparsity properties of the cell-cell interaction graph.

Given a cell-cell interaction graph $\mathcal{G} = (V, E)$, the set $V$ consist of $N$ nodes, each representing a cell $\{v_1, v_2, \ldots, v_N\}$ with $E$ forming the edges that define cellular interactions captured in the adjacency matrix $A \in \mathbb{R}^{N \times N}$. Each node $v_i$ has an associated gene expression profile represented as feature vector $x_i$ in the feature matrix $X \in \mathbb{R}^{N \times d}$. Nodes are labeled with their cell type $Y=\{y_1, y_2,\ldots, y_N\}$, serving as training data for the node classification task. We developed a two-layer Graph Convolutional Network (GCN), which effectively maps the node embeddings based on cell type information $Y$. This model generates a new graph $\mathcal{G}'$ with an adjusted adjacency matrix $S$ designed to achieve three key objectives: (i) it encapsulates cell type information; (ii) it preserves essential graph properties such as low rank and sparsity; and (iii) it maintains feature smoothness across neighboring nodes. Specifically, the GCN is structured by the function $f_\theta(X,A) = \text{softmax}(\hat{A}\sigma(\hat{A}XW_1)W_2)$ ,where $\hat{A} = \widehat{D}^{-1/2}(A+I)\widehat{D}^{-1/2}$ and $\widehat{D}$ is the diagonal matrix of $A+I$ with $\widehat{D}_{ii} = 1 + \sum_j A_{ij}$. $\sigma$ represents a *ReLU* activation function. To achieve objective (i), one object function of the GCN can be formulated as:

$$\min_\theta \mathcal{L}_{Cls}(\theta, A, X, Y) = \sum_{v_i \in V} \text{cross-entropy}(f_\theta(X,A)_i, y_i) \quad (1)$$

To enable the GCN to effectively learn the symmetric adjacency matrix $S$ (initialized as $A$) of the newly formulated undirected graph $\mathcal{G}'$ and to achieve objective (ii), which entails maintaining essential graph properties of low rank and sparsity, an appropriately designed loss function was incorporated into the model's training:

$$\operatorname*{arg\,min}_{S\in[0,1]^{N\times N}} \mathcal{L}_{graph} = \|A - S\|_F^2 + R(S), \quad s.t.,\ S = S^T \quad (2)$$

$R(S)$ represents the constraints applied to $S$ to ensure the preservation of properties of low rank and sparsity. According to [26], it is established that the minimization of the $\ell_1$ norm and the nuclear norm of a matrix promotes sparsity and low rank characteristics, respectively. Consequently, Equation (2) can be reformulated as follow:

$$\operatorname*{arg\,min}_{S\in[0,1]^{N\times N}} \mathcal{L}_{graph} = \|A - S\|_F^2 + w_s\|S\|_1 + w_l\|S\|_*, s.t.,\ S = S^T, \quad (3)$$

where $w_s$ and $w_l$ are predetermined parameters that modulate the influence of sparsity and low rank properties, respectively. In this work, the values of $\alpha$ and $\beta$ were respectively set at 5e-4 and 1.5. The third loss function for objective aims to maintain node feature smoothness in the learned graph, which is represented by the following term $\mathcal{L}_{node}$

$$\mathcal{L}_{node} = tr(X^T L X), \tag{4}$$

where $L$ is the normalized graph Laplacian matrix of $S$. The final objective function is as follow:

$$\arg\min_{S,\theta} \mathcal{L} = w_c \mathcal{L}_{Cls} + \mathcal{L}_{graph} + w_n \mathcal{L}_{node}, \tag{5}$$

$w_c$ and $w_n$ were set as 0.1 and 1. To solve this loss, we employ an alternating optimization schema (details provided in Supplementary File 1) to jointly update $\theta$ and $S$.

## 2.4 Cell type annotation with graph domain adaptation

As an increasing number of GNN algorithms are developed for cell type prediction using scRNA-seq data, the diversity in the construction of these graphs—whether cell-cell, gene-gene, or cell-gene graphs—and their formulation methods continue to expand. Despite their potential, these methods often face generalization challenges, especially when scRNA-seq data used for inference comes from patients with different health conditions, or when the data is collected using varied methodologies across different studies. This variability can lead to unique interconnection patterns in the data. In the context of GNNs used for single-cell prediction in cell-cell graphs, transfer issues in node classification (i.e., cell type annotation) become apparent, particularly when annotating cell types within the TME of one patient based on a model trained with scRNA-seq data from another. This highlights the inherent biological and contextual differences between individuals. In this work, we introduce for the first time the concept of graph domain adaptation (GDA) to address the challenges associated with applying a learned cell-cell graph for cell type annotation across different individuals.

Specifically, we train a model to predict cell types from reference scRNA-seq data $\mathcal{G}_S = (\mathcal{V}_S, \mathcal{E}_S, \mathcal{X}_s)$ which comprises a feature encoder, $\phi: \chi \to \mathcal{H}$ and a classifier, $g: \mathcal{H} \to Y$. In this work, $\phi$ implemented as a two-layer GCN, similar to the GSL model, and $g$ a two-layer Multi-Layer Perceptron. Given the query data $\mathcal{G}_T = (\mathcal{V}_T, \mathcal{E}_T, \mathcal{X}_T)$, the objective is to use GDA on this model to minimize the target error $\varepsilon_T(g \circ \phi)$.

In GDA for graphs with the adjacency matrix $A$ and node labels $Y$, shifts in graph structure, indicated by $\mathbb{P}_S(A, Y) \neq \mathbb{P}_T(A, Y)$, typically manifest as either conditional structure shift (CSS), label shift (LS), or a combination of both. With decomposition as $\mathbb{P}(A, Y) = \mathbb{P}(A|Y)\mathbb{P}(Y)$, CSS and LS are defined as $\mathbb{P}_S(A|Y) \neq \mathbb{P}_T(A|Y)$ and $\mathbb{P}_S(Y) \neq \mathbb{P}_T(Y)$, respectively. To address these, the GDA framework of Pairwise Alignment (PA) is employed. Specifically, PA constructs the matrix of density ratio between the target and source graphs $\gamma \in \mathbb{R}^{|\mathcal{Y}| \times |\mathcal{Y}|}$ to reweight edges for GNN encoding, which can be formulated as follow:

$$[\gamma]_{i,j} = \frac{\mathbb{P}_T(Y_v = j | Y_u = i, v \in \mathcal{N}_u)}{\mathbb{P}_S(Y_v = j | Y_u = i, v \in \mathcal{N}_u)}, \ \forall i, j \in \mathcal{Y}. \tag{6}$$

$[\gamma]_{i,j}$ denotes the density ratio from class-$i$ nodes to class-$j$ nodes between two graphs. To optimize γ, PA further decompose it into another two weights, $w \in \mathbb{R}^{|\mathcal{Y}| \times |\mathcal{Y}|}$ and $\alpha \in \mathbb{R}^{|\mathcal{Y}| \times 1}$, which are defined as:

$$[w]_{i,j} = \frac{\mathbb{P}_T(Y_u = i, Y_v = j | e_{uv} \in \mathcal{E}_T)}{\mathbb{P}_S(Y_u = i, Y_v = j | e_{uv} \in \mathcal{E}_S)}, \tag{7}$$

$$[\alpha]_i = \frac{\mathbb{P}_T(Y_u = i | e_{uv} \in \mathcal{E}_T)}{\mathbb{P}_S(Y_u = i | e_{uv} \in \mathcal{E}_S)}, \ \forall i, j \in \mathcal{Y}. \tag{8}$$

Given $w$, $\alpha$ can be derived via:

$$[\alpha]_i = \frac{\sum_{j \in \mathcal{Y}} ([w]_{i,j} \mathbb{P}_S(Y_u = i, Y_v = j | e_{uv} \in \mathcal{E}_S))}{\mathbb{P}_S(Y_u = i | e_{uv} \in \mathcal{E}_S)}. \tag{9}$$

Given w and $\alpha$, $\gamma$ can then be estimated via:

$$\gamma = \text{diag}(\alpha)^{-1} w. \quad (10)$$

To estimate *w*, PA uses pair-wise pseudo-label distribution alignment using the matrix Σ, which represents the joint distribution of predicted and true edge types, and the vector *v*, the distribution of predicted edge types in the target domain. Based on soft label of node *u* yielded from the classifier *g*, Σ and *v* can be estimated via:

$$\left[\hat{\Sigma}\right]_{i,j,i',j'} = \frac{1}{|\mathcal{E}_S|} \sum_{e_{uv} \in \mathcal{E}_S, y_u = i', y_v = j'} [g(h_u^{(L)})]_i \times [g(h_v^{(L)})]_j, \quad (11)$$

$$[\hat{v}]_{ij} = \frac{1}{|\mathcal{E}_T|} \sum_{e_{u'v'} \in \mathcal{E}_T} [g(h_{u'}^{(L)})]_i \times [g(h_{v'}^{(L)})]_j. \quad (12)$$

Finally, *w* can be solved via:

$$\min_w \left\| \hat{\Sigma} w - \hat{v} \right\|_2, \text{ s.t. } w \geq 0, \text{and } \sum_{i,j} [w]_{i,j} \mathbb{P}_S(Y_u = i, Y_v = j | e_{uv} \in \mathcal{E}_S) = 1. \quad (13)$$

PA addresses CSS through such an iterative method. It starts by using an estimated γ as edge weights in the source graph to reduce the difference between $\mathbb{P}_S(H^{(L)}|Y)$ and $\mathbb{P}_T(H^{(L)}|Y)$. With this reduced difference, *w* can be estimated more accurately, leading to better refinement of γ. Continuous iterations allow γ to progressively improve the conditional alignment, effectively tackling CSS.

To tackle LS, PA calculates the ratio between the source and target label distributions by aligning the node-level pseudo-label distributions, $\beta \in \mathbb{R}^{|\mathcal{Y}| \times 1}$, in which $[\beta]_i = \frac{\mathbb{P}_T(Y=i)}{\mathbb{P}_S(Y=i)}, \forall i \in \mathcal{Y}$. Similar to the solution for estimating *w*, *β* is determined by resolving the linear equation *μ*=C*β*, where C represents the confusion matrix from the classifier in the source domain and *μ* denotes the distribution of the predicted labels in the target domain. Based on soft label of node *u* yielded from the classifier *g*, C and *μ* are estimated as

$$[\hat{\mathrm{C}}]_{i,i'} = \frac{1}{|\mathcal{V}_S|} \sum_{u \in \mathcal{V}_S, y_u = i'} [g(h_u^{(L)})]_i \quad (14)$$

$$[\hat{\mu}]_i = \frac{1}{|\mathcal{V}_T|} \sum_{u' \in \mathcal{V}_T} [g(h_u^{(L)})]_i \quad (15)$$

Ultimately, β can be determined by solving a least squares problem with constraints that ensure a valid target label distribution:

$$\min_{\beta}\left\|\hat{C}\beta-\hat{\mu}\right\|_2 \quad s.t.\ \beta\geq 0,\ \ \sum_i[\beta]_i\mathbb{P}_S(Y=i)=1 \tag{16}$$

The feature encoder $\phi$ and classifier $g$ are trained via the weighted cross-entropy loss:

$$\mathcal{L}_C^{\beta}=\frac{1}{|\mathcal{V}_S|}\sum_{v\in\mathcal{V}_S}[\beta]_{y_v}\text{cross-entropy}(y_v,\hat{y}_v) \tag{17}$$

The entire algorithm of PA for addressing the GDA problem in a reference-query pair of cell-cell graphs is shown in Algorithm 1.

**Algorithm 1.** GDA for reference-query cell-cell graphs

1: **Input** The reference graph $\mathcal{G}_S$ with cell type labels $Y_S$; The query graph $\mathcal{G}_T$; The feature encoder $\phi$ and classifier $g$; The numbers of epoch and period for training, $n$ and $t$.
2: Initialize $w$, γ and $\beta$,
3: **while** epoch < $n$ or not converged **do**
4: Use γ to add weights to reference cell-cell graph
5: Get predicted cell labels of reference graph through $\hat{Y}_S=g(\phi(x_S,A_S))$
6: Update $\phi$ and $g$ using Eq.(17)
7: **if** epoch ==0(mod $t$) **then**
8: Get predicted cell labels of two graphs, $\hat{Y}_S$ and $\hat{Y}_T=g(\phi(x_T,A_T))$
9: Update $\hat{\Sigma},\hat{v},\hat{C}$ and $\hat{\mu}$ using Eq.(11-12) and Eq.(14-15)
10: Optimize for $w$ using Eq.(13) and calculate γ using Eq.(10)
11: Optimize for $\beta$ using Eq.(16)
12: **end if**
13: **end while**

# 3 Results

## 3.1 Predictive Performance of the scGSL Model Across Three Disease Types at the Patient Level

To assess the accuracy of the scGSL model in cell type identification, we employed it to predict paired cell types across different patient samples for three distinct diseases. In each dataset, every pair of patient samples acted as both "training set" and "test set" in a comprehensive combinatorial setup. Each experiment was evaluated based on four key metrics: Accuracy (ACC), Precision (PRE), Recall (REC), and F1 Score (F1). Fig. 3 presents the overall experimental results in the form of violin plots, where each subplot displays outcomes based on the aggregate dataset and the three disease-specific datasets. Each point within the plots represents an experimental pair of patient samples, providing a detailed visual representation of the model's performance across various conditions.

Overall, the model achieved an average accuracy of 84.83%, with precision at 86.23%, recall at 81.51%, and an F1 score of 80.92% across all datasets, indicating robust predictive capabilities. In assessing the scGSL model across three distinctive datasets, each showed varied but generally effective performance metrics. The Leukemia dataset showed a robust average accuracy of 86.85% and precision at 78.46%, though it had slightly lower recall and F1 scores at 75.23% and 73.39%, respectively. The Breast Invasive Carcinoma dataset maintained a balanced performance with an average accuracy of 83.56% and a notably high precision of 89.42%. In the Colorectal Cancer dataset, the model excelled with an average accuracy of 86.52%, complemented by consistent precision and recall rates of 86.44% and 86.07%. These findings across the datasets underline the scGSL model's robust predictive capabilities and highlight its potential applicability in diverse clinical scenarios.

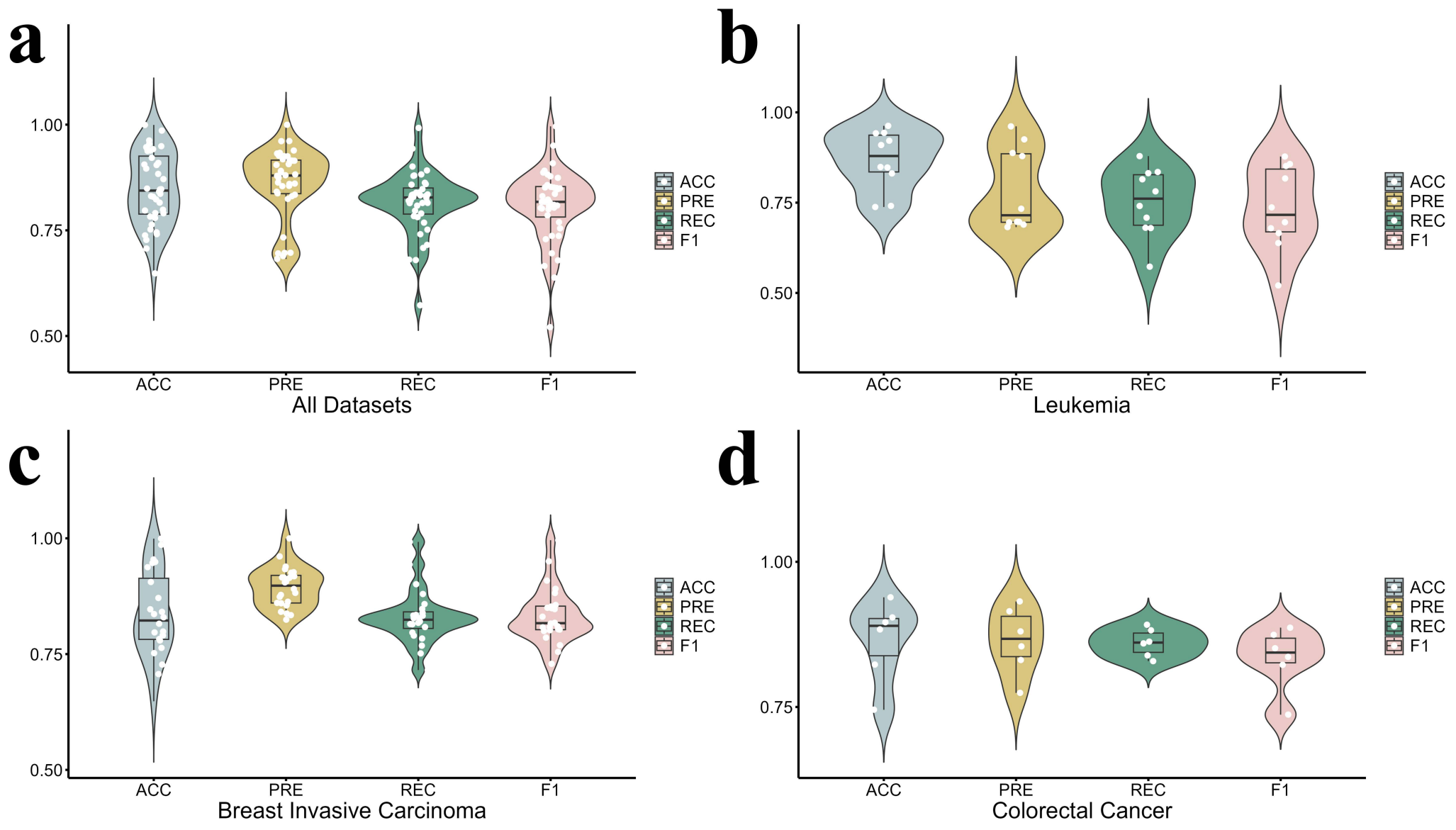

**Fig. 3** Violin Scatter Plots of scGSL Performance in Patient Pair Experiments.

## 3.2 Comparison of scGSL Model Performance with Other Models

In the field of single-cell RNA sequencing, several computational methods have been developed for cell type identification, each utilizing different strategies. These include marker-based methods like MarkerCount [27] and CHETAH [28], which rely on known genetic markers; supervised machine learning approaches such as singleCellNet [29], scLearn [30], and scPred [31], which use labeled datasets to train predictive models; reference-based mapping techniques like scmapCell [32], SingleR, and sciBet [33], which match query cells to a reference database; and comprehensive frameworks like Seurat and scClassify [34]. To thoroughly evaluate the efficacy of our scGSL model, we have conducted comparative experiments using these ten established methods across all available datasets. The results of these comparisons are meticulously presented in Table 1 for the Leukemia dataset, Table 2 for the Breast Invasive Carcinoma dataset, and Table 3 for the Colorectal Cancer dataset. Additionally, a partial comparison of the embeddings learned by these methods is visually displayed in Fig. 4.

In the comparative evaluation across three datasets, our scGSL model consistently demonstrated superior performance. On the Leukemia dataset, it achieved the highest average accuracy (ACC) of 86.85%, outperforming the second-best SingleR model by 3.07%. In the Breast Invasive Carcinoma dataset, scGSL led with an average ACC of 82.14%, surpassing SingleR by 3.20%. Similarly, on the Colorectal Cancer dataset, it recorded the highest average ACC of 86.52%, exceeding the second-best scLearn by 2.93%. These results highlight scGSL's robust ability to accurately classify cell types, outshining other methods in various cancer contexts.

Table 1. Comparative Performance of Different Models on the Leukemia Dataset (Best Performance in Bold, Second Best Underlined).

| Experiment | Seurat | SingleR | CHETAH | scmapCell | singleCellNet | scLearn | scPred | Marker Count | scClassify | sciBet | Ours |
|---|---|---|---|---|---|---|---|---|---|---|---|
| AYL050-OX1164 | **88.49** | 81.98 | 67.59 | 75.21 | 81.12 | 79.00 | 73.45 | 65.42 | 77.89 | 79.20 | 83.09 |
| OX1164-AYL050 | 87.58 | 87.38 | 84.45 | 69.15 | 84.75 | 85.25 | 86.39 | 76.06 | 87.43 | 86.24 | **92.15** |
| PBMMC_1-PBMMC_2 | 91.71 | 83.88 | 80.91 | 66.85 | 88.78 | 92.76 | **96.50** | 49.93 | 90.98 | 75.84 | 94.20 |
| PBMMC_2-PBMMC_1 | 88.38 | 81.99 | 76.84 | 67.35 | 85.09 | 82.79 | **92.46** | 79.15 | 72.58 | 72.67 | 90.95 |
| PBMMC_1-PBMMC_3 | 78.84 | 83.73 | 83.27 | 55.58 | 77.83 | 84.32 | 84.55 | 83.23 | 74.31 | 84.28 | **84.87** |
| PBMMC_3-PBMMC_1 | 70.72 | **91.93** | 85.45 | 61.76 | 83.67 | 84.56 | 73.11 | 70.81 | 66.90 | 84.38 | 84.74 |
| PBMMC_2-PBMMC_3 | 94.24 | 88.25 | 85.37 | 69.01 | 86.79 | 88.89 | 75.23 | 85.19 | 82.95 | 84.78 | **94.33** |
| PBMMC_3-PBMMC_2 | 95.84 | 96.12 | 85.38 | 70.21 | 95.38 | 96.19 | **97.13** | 93.67 | 95.66 | 94.27 | 96.26 |
| P6-P7 | 62.98 | **74.37** | 72.01 | 23.11 | 71.28 | 73.47 | 66.23 | 67.53 | 72.09 | 73.23 | 73.80 |
| P7-P6 | 67.60 | 68.14 | **90.79** | 19.58 | 69.04 | 61.73 | 69.40 | 73.19 | 67.24 | 67.69 | 74.10 |
| Average | 82.64 | 83.78 | 81.21 | 57.78 | 82.37 | 82.90 | 81.45 | 74.42 | 78.80 | 80.26 | **86.85** |

Table 2. Comparative Performance of Different Models on the Breast Invasive Carcinoma Dataset (Best Performance in Bold, Second Best Underlined).

| Experiment | Seurat | SingleR | CHETAH | scmapCell | singleCellNet | scLearn | scPred | Marker Count | scClassify | sciBet | Ours |
|---|---|---|---|---|---|---|---|---|---|---|---|
| tnbc1-tnbc6k | 98.36 | **99.67** | 76.31 | 98.46 | 94.78 | 78.62 | 97.86 | 75.38 | 98.03 | 89.43 | 98.59 |
| tnbc6k-tnbc1 | 99.86 | 99.86 | 99.64 | 99.35 | **99.93** | **99.93** | 99.42 | 99.49 | **99.93** | 98.56 | **99.93** |
| sc5rJUQ024-sc5rJUQ026 | 48.01 | 67.82 | 54.20 | 28.75 | 71.84 | 54.83 | 55.28 | 49.59 | 56.32 | 53.20 | **77.93** |
| sc5rJUQ026-sc5rJUQ024 | 73.91 | 81.29 | 69.09 | 46.23 | 75.74 | 53.97 | 65.12 | 48.66 | 67.66 | 67.98 | **81.55** |
| sc5rJUQ024-sc5rJUQ033 | 65.57 | **83.78** | 54.99 | 41.68 | 69.70 | 69.49 | 44.15 | 46.39 | 65.05 | 59.67 | 70.69 |
| sc5rJUQ033-sc5rJUQ024 | 73.76 | 77.58 | 53.39 | 13.84 | 73.61 | 71.37 | **79.68** | 62.70 | 72.94 | 51.90 | 79.07 |
| sc5rJUQ024-sc5rJUQ050 | 90.79 | 93.47 | 84.13 | 70.61 | 90.99 | 90.49 | 71.57 | 86.22 | 85.55 | 84.92 | **95.43** |
| sc5rJUQ050-sc5rJUQ024 | 73.61 | 76.18 | 63.05 | 45.15 | 76.56 | 71.48 | 55.40 | 67.78 | 71.28 | 65.24 | **79.68** |
| sc5rJUQ024-sc5rJUQ060 | 68.08 | 79.59 | 66.36 | 24.81 | 78.67 | 58.39 | 57.78 | 45.47 | 53.03 | 61.70 | **83.64** |
| sc5rJUQ060-sc5rJUQ024 | 75.45 | 80.04 | 65.21 | 31.35 | 78.28 | 61.88 | 76.09 | 62.52 | 67.83 | 61.30 | **84.68** |
| sc5rJUQ026-sc5rJUQ033 | 54.86 | 47.87 | 57.90 | 46.52 | 63.25 | 53.04 | 56.76 | 41.61 | 56.91 | 54.99 | **64.81** |
| sc5rJUQ033-sc5rJUQ026 | 49.73 | 80.14 | 52.35 | 23.74 | 73.87 | 60.47 | 44.77 | 63.54 | 67.96 | 29.65 | **82.94** |
| sc5rJUQ026-sc5rJUQ050 | 91.52 | 94.86 | 90.56 | 81.18 | 92.68 | 90.62 | 80.52 | 84.69 | 84.00 | 86.81 | **94.90** |
| sc5rJUQ050-sc5rJUQ026 | 65.93 | **80.91** | 57.31 | 35.15 | 74.23 | 70.44 | 57.18 | 65.03 | 65.30 | 55.32 | 79.87 |
| sc5rJUQ033-sc5rJUQ050 | 81.58 | 85.49 | 33.83 | 49.04 | 90.56 | 89.56 | 89.36 | 89.46 | 89.70 | 74.22 | **93.77** |
| sc5rJUQ050-sc5rJUQ033 | 71.44 | 55.87 | 52.81 | 38.90 | 63.51 | 73.99 | 66.29 | 38.98 | 66.27 | 66.87 | **79.03** |
| sc5rJUQ026-sc5rJUQ060 | 65.56 | **83.04** | 71.84 | 29.88 | 71.33 | 53.76 | 59.25 | 36.96 | 67.16 | 50.32 | 72.74 |
| sc5rJUQ060-sc5rJUQ026 | 49.10 | 73.96 | 43.32 | 24.32 | 69.04 | 54.15 | 66.74 | 60.83 | 60.97 | 35.97 | **78.25** |
| sc5rJUQ033-sc5rJUQ060 | 69.42 | 79.02 | 74.49 | 14.67 | 78.70 | 73.47 | 60.49 | 57.43 | 70.60 | 63.42 | **84.12** |
| sc5rJUQ060-sc5rJUQ033 | 71.26 | 71.52 | 52.34 | 41.53 | 74.77 | 70.92 | **79.55** | 44.26 | 72.51 | 66.50 | 75.21 |
| sc5rJUQ050-sc5rJUQ060 | 69.77 | 76.15 | 64.48 | 25.38 | 74.81 | 69.90 | 72.64 | 60.40 | 62.21 | 64.45 | **76.34** |
| sc5rJUQ060-sc5rJUQ050 | 86.75 | 93.67 | 51.33 | 47.75 | 85.32 | 88.44 | 60.30 | 84.63 | 87.87 | 76.97 | **94.67** |
| GSM4555888-GSM4555891 | 84.35 | **91.96** | 72.09 | 68.04 | 85.69 | 72.40 | 73.97 | 85.45 | 62.47 | 84.82 | 87.12 |
| GSM4555891-GSM4555888 | 90.15 | 82.36 | 87.13 | 29.19 | 89.41 | 89.31 | 63.50 | 83.23 | 90.41 | 89.41 | **90.57** |
| Average | 71.39 | 78.94 | 62.37 | 38.99 | 77.39 | 70.11 | 65.29 | 62.08 | 70.18 | 63.89 | **82.14** |

Table 3. Comparative Performance of Different Models on the Colorectal Cancer Dataset (Best Performance in Bold, Second Best Underlined).

| Experiment | Seurat | SingleR | CHETAH | scmapCell | singleCellNet | scLearn | scPred | MarkerCount | scClassify | sciBet | Ours |
|---|---|---|---|---|---|---|---|---|---|---|---|
| scrEXT009-scrEXT010 | 76.45 | 85.20 | 79.12 | 54.36 | 89.04 | 84.13 | 64.81 | 76.39 | 87.10 | 85.66 | **89.61** |
| scrEXT010-scrEXT009 | 66.43 | 83.95 | 76.93 | 40.11 | 79.11 | 87.50 | 75.58 | 84.88 | 84.97 | 81.21 | **88.35** |
| scrEXT009-scrEXT020 | 62.14 | 90.03 | 55.85 | 49.50 | 73.95 | 85.46 | 26.47 | 69.38 | 78.16 | 85.99 | **90.45** |
| scrEXT020-scrEXT009 | 71.48 | 77.57 | 72.69 | 29.87 | 70.13 | 73.26 | **77.78** | 68.44 | 70.55 | 73.02 | 74.53 |
| scrEXT010-scrEXT020 | 82.08 | 92.52 | 67.77 | 57.86 | 82.38 | 89.97 | 90.68 | 75.07 | 91.04 | 89.97 | **93.89** |
| scrEXT020-scrEXT010 | 70.77 | 64.07 | 80.94 | 32.28 | 80.56 | 81.23 | **86.41** | 75.32 | 80.51 | 82.21 | 82.32 |
| Average | 71.56 | 82.22 | 72.22 | 44.00 | 79.19 | 83.59 | 70.29 | 74.92 | 82.06 | 83.01 | **86.52** |

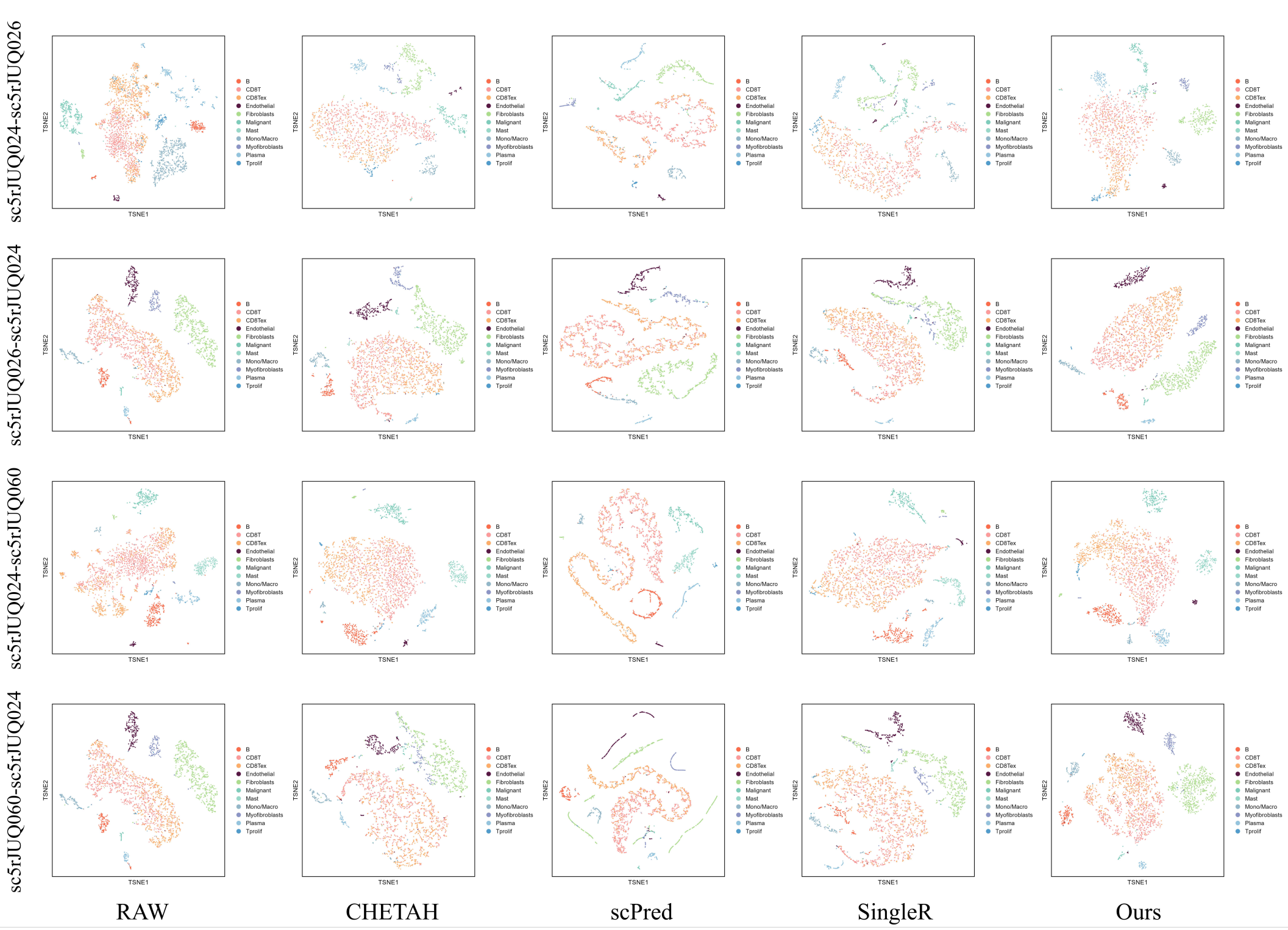

**Fig. 4** Comparison of Cell Embeddings and Raw Data Distributions between scGSL and Other Methods.

## 3.3 Ablation Study for scGSL with Different Settings

In constructing the cell-cell graph for the scGSL model, HVGs are first selected, followed by the application of the KNN algorithm to establish an initial graph structure. Given that scGSL is a model based on GNN, this section systematically examines the impact of both the selection of the KNN algorithm and the HVG filtering criteria on the predictive performance of scGSL.

For the former, Fig. 5a displays violin scatter plots comparing the scGSL model's performance across different datasets under various KNN settings. The standard setting of K=0.2% total cell count and embedding size of 128 (K=0.2%, ES=128) consistently delivered superior results, achieving the highest average accuracy and F1-score. Other tested settings included K=0.1% & ES=128, K=0.5% & ES=128, K=0.2% & ES=64, and K=0.2% & ES=256. None of these configurations could match the ACC and F1 results of the standard setting. This emphasizes the effectiveness and robustness of the K=0.2% and ES=128 configuration in the scGSL model for optimal cell type classification performance.

To compare different gene selection strategies, we collated gene sets from six gene signaling pathway databases, including KEGG and Reactome, as well as the cell state marker gene database (CancerSEA database) and the LRI database [19, 35-39]. We used the genes found within each of these individual databases as criteria for gene selection, which were then used to construct node features in our graph-based model. Fig. 5b showcases a heatmap of the scGSL model's performance, combining these varied selection approaches on the E-MTAB-8107 dataset, which includes data from 10 pairs of patients across five distinct patients. In our comparative analysis of gene selection strategies across several databases, our method (HVG3000) demonstrated superior performance with an ACC of 81.47%, PRE of 88.39%, REC of 82.49%, and F1 of 82.58%. Fig. 5c illustrates a comparison of cell type prediction accuracies in the GSE132509 dataset, using the scGSL model both with and without the GSL component. The results demonstrate that the inclusion of GSL significantly enhances predictive accuracy. Specifically, the average accuracy with GSL was 90.89%, compared to 87.55% when GSL was not utilized. This improvement underscores the effectiveness of GSL in enhancing the precision of cell type predictions in complex biological datasets.

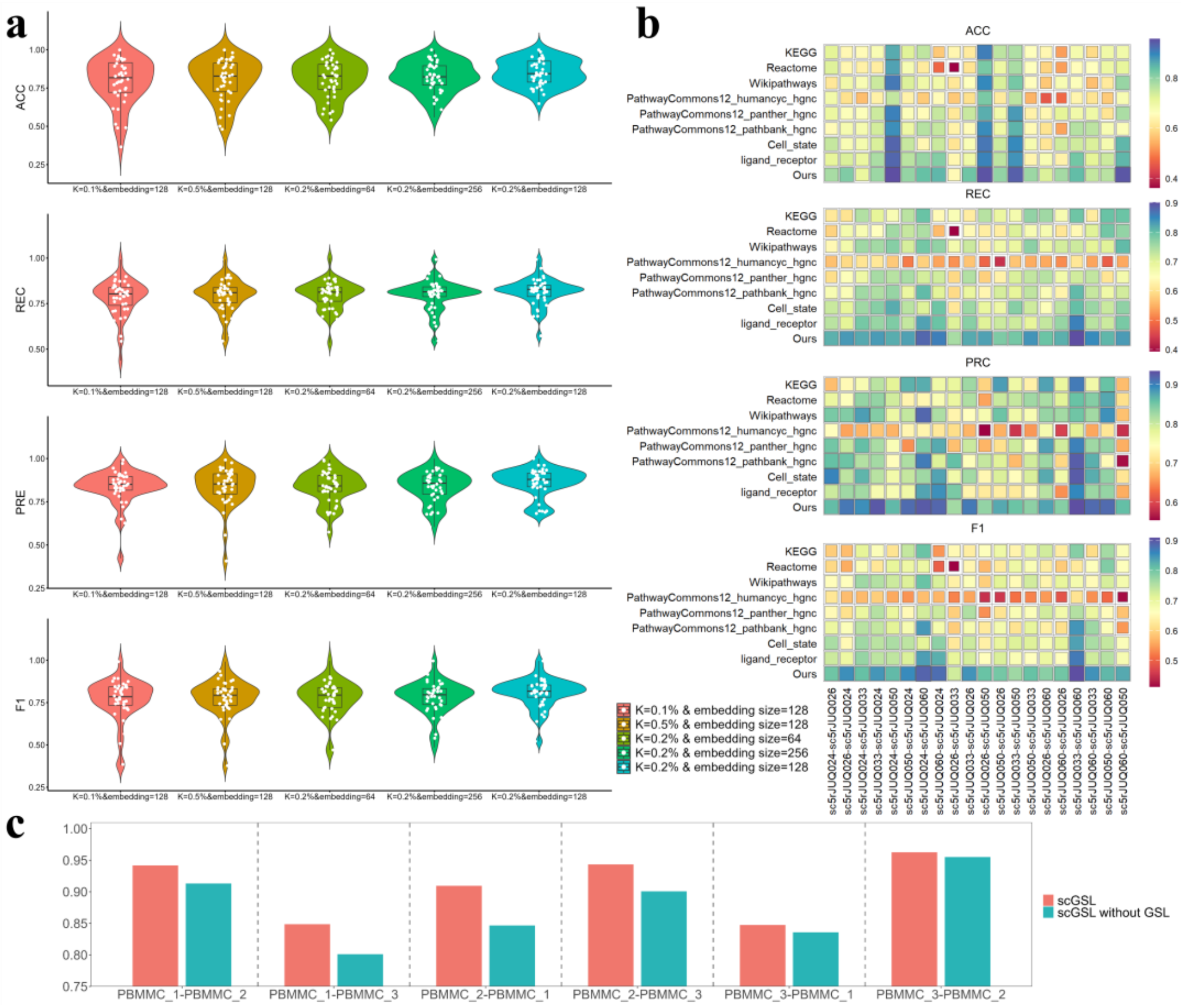

**Fig. 5** Performance Comparison of scGSL under Various Settings. a) Comparison of performance metrics using varying sparsity levels and neighbor counts during cell-cell graph construction. b) Performance differences when employing various pathway databases for gene filtering. c) Comparative analysis of performance with and without GSL.

### 3.4 Evaluating the Interpretability of Cell-Cell Graphs Generat-ed from scGSL in TME

In this section, we explore the interpretability of disease-specific cell-cell graphs derived from scGSL by reviewing existing literature. Specifically, we focus on gene pairs involved in cell-cell interactions within the microenvironments of three diseases: Leukemia, Breast Invasive Carcinoma, and Colorectal Cancer. These gene pairs have been confirmed in prior studies as significant for their roles in intercellular communication within these diseases. Utilizing these validated gene pairs, we assess the statistical significance of these interactions in the constructed cell-cell graphs (visualized in Fig. 6). In the Leukemia dataset, the "SPRY1-LAT" and "BSG-ATP2B4" gene pairs within CD8T cells are investigated, confirmed by studies referenced in PMID19915061 [40] and PMID26729804 [41]; in the Breast Invasive Carcinoma dataset, the "ATP2A3-SP1" and "ZAP70-TLN1" gene pairs within CD8T cells are investigated, as established in the literature with PMIDs 22851172 [42] and 20488542 [43]; in the Colorectal Cancer dataset, the "BIRC5-CASP6" and "PSMC3-SIRT7" gene pairs within malignant cells are investigated, documented in PMIDs 23856250 [44] and 28435470 [45].

Specifically, we focused on the above particular gene pairs within a certain cell type, such as the CD8T cell subgraph in the Leukemia Dataset. We categorized the cells into two groups: one containing cells connected by an edge and the other consisting of cells without an edge, which served as the background group. We then conducted the Mann-Whitney U test to compare the expression levels of the two genes in the gene pair between these groups. If the p-value from this test is less than 0.05, it indicates that the differences in gene expression are statistically significant. Such a significant result would support the claim that the gene pair's expression differences, as learned by the scGSL from the cell-cell graph, are biologically meaningful.

It is noted that the generation of the cell-cell graphs was generated in an unsupervised manner, meaning that the gene pairs validated were not involved in the model training process. As a result, Tables 4, 5, and 6 present findings from three different cancer datasets: Leukemia, Breast Invasive Carcinoma, and Colorectal Cancer. In the Leukemia dataset, the gene pair "SPRY1-LAT" achieved significant p-values (less than 0.05) in 7 out of 12 experiments, while "BSG-ATP2B4" showed significance in 10 out of 12 experiments. In the dataset for Breast Invasive Carcinoma, every experiment involving the gene pair "ATP2A3-SP1" resulted in significant p-values, and "ZAP70-TLN1" was significant in 6 out of 8 experiments. In the Colorectal Cancer dataset, "BIRC5-CASP6" was significant in 7 out of 8 experiments, and "PSMC3-SIRT7" in 6 out of 8. These results underscore the robustness and biological relevance of the unsupervised cell-cell graph approach, as it successfully identified biologically meaningful gene interactions across various types of cancer without prior knowledge of these interactions during the model's training phase.

Table 4. Statistical Significance of Known Gene Interactions for CD8+ T Cells in Constructed Cell-Cell Graphs on the Leukemia Dataset (P-Value < 0.05 Highlighted).

| Experiment | SPRY1 | LAT | BSG | ATP2B4 |
|---|---|---|---|---|
| PBMMC_1-2 | **1.67403e-04** | **9.46e-27** | **7.01e-33** | **7.92883e-04** |
| PBMMC_2-1 | 1.15814e-01 | 3.42939e-01 | **3.92324e-03** | **2.50337e-02** |
| PBMMC_1-3 | **1.53503e-03** | **8.07e-05** | **6.01e-05** | **7.62435e-03** |
| PBMMC_3-1 | 1.22120e-01 | 1.46373e-01 | **3.58655e-02** | **9.62522e-03** |
| PBMMC_2-3 | **4.06214e-03** | **4.93204e-03** | **1.51e-06** | **1.42773e-02** |
| PBMMC_3-2 | 3.44669e-01 | **5.62577e-03** | 3.48405e-01 | 1.68556e-01 |

Table 5. Statistical Significance of Known Gene Interactions for CD8+ T Cells in Constructed Cell-Cell Graphs on the Breast Invasive Carcinoma Dataset (P-Value < 0.05 Highlighted).

| Experiment | ATP2A3 | SP1 | ZAP70 | TLN1 |
|---|---|---|---|---|
| sc5rJUQ024-26 | **1.29373e-04** | **1.27737e-04** | 1.17708e-01 | **3.21875e-02** |
| sc5rJUQ026-24 | **2.18143e-02** | **6.93e-09** | 1.76564e-01 | **4.67e-27** |
| sc5rJUQ024-33 | **1.29e-12** | **1.04e-07** | **7.00e-07** | **8.78e-16** |
| sc5rJUQ033-24 | **1.77e-13** | **2.57e-15** | **8.95e-07** | **4.98e-21** |

Table 6. Statistical Significance of Known Gene Interactions for Malignant Cells in Constructed Cell-Cell Graphs on the Colorectal Cancer Dataset (P-Value < 0.05 Highlighted).

| Experiment | BIRC5 | CASP6 | PSMC3 | SIRT7 |
|---|---|---|---|---|
| scrEXT009-10 | **2.41032e-02** | **7.15088e-03** | **9.11823e-03** | 4.93301e-01 |
| scrEXT010-09 | **4.35e-16** | **1.11573e-03** | **1.16e-12** | **1.68677e-02** |
| scrEXT009-20 | **1.19508e-02** | 3.99906e-01 | **3.38843e-01** | 9.50344e-02 |
| scrEXT020-09 | **6.53e-15** | **1.97e-05** | **6.16e-15** | **7.71e-06** |

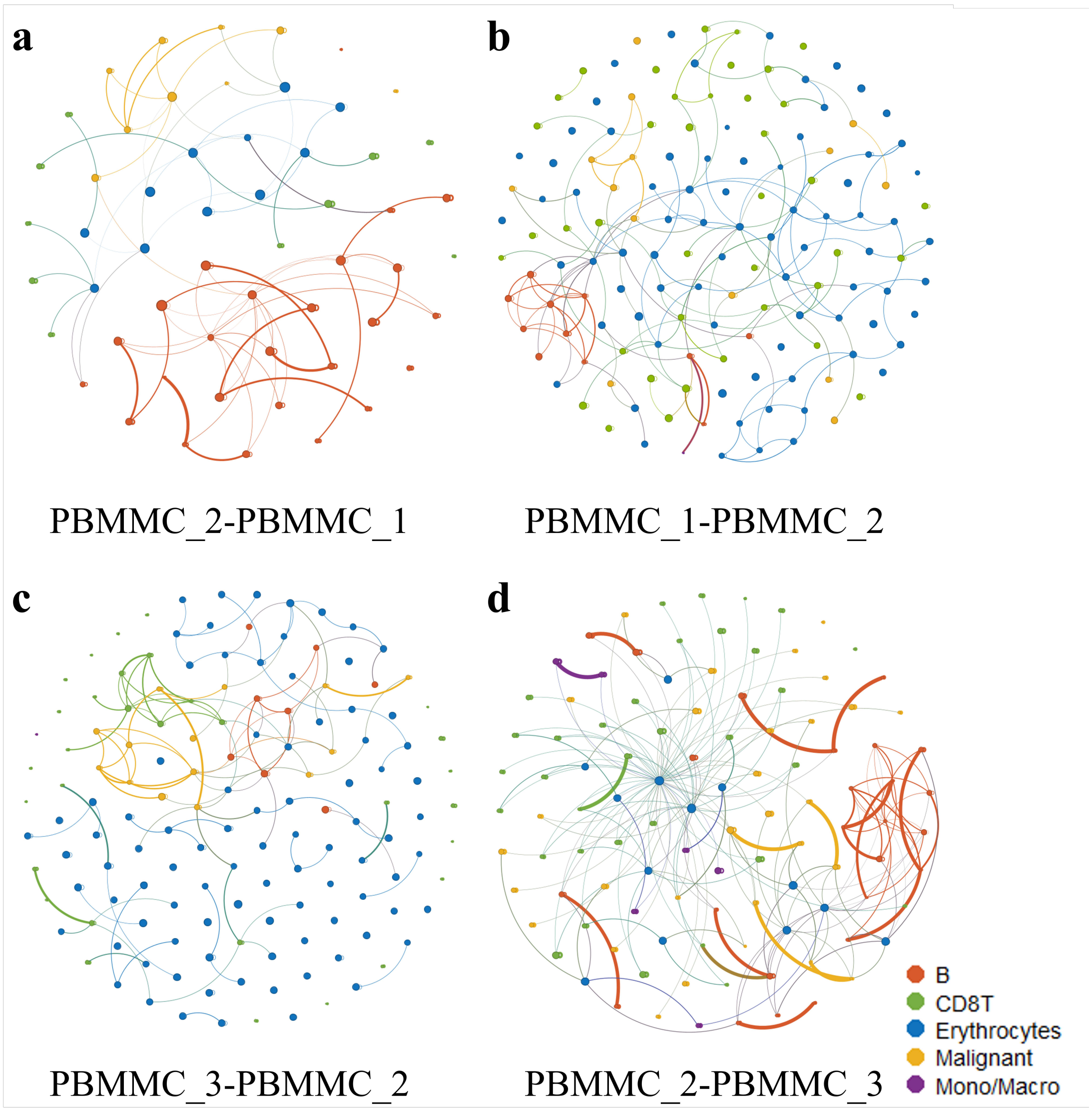

**Fig. 6** Visualization of Cell-Cell Graphs Derived from scGSL on the Leukemia Dataset.

## 4 Conclusions

The scGSL model represents a significant advancement in the analysis of non-spatial transcriptomic scRNA-seq data within the context of cancer research. By leveraging GNN and graph domain adaptation, our model not only enhances the accuracy of cell type annotation but also provides a deeper understanding of the intricate cellular interactions within the TME. The ability of the scGSL model to autonomously generate and refine cell-cell interaction graphs based on intrinsic labels ensures a more biologically accurate representation of intercellular communication. This is crucial for identifying potential therapeutic targets and understanding the complex network of interactions that drive tumor progression and resistance to therapies.

Despite its robust performance, the scGSL model faces certain challenges and limitations. One significant challenge is the inherent complexity of TMEs, which can vary widely between different types of cancers and even between patients with the same type of cancer. This variability can affect the generalizability of the model across different datasets and cancer types. Additionally, the reliance on labeled data from specific cancer types for training may limit the model’s applicability to cancers with less available data. Furthermore, the model's performance is contingent on the quality and completeness of the underlying scRNA-seq data, which can be affected by technical variations and sampling errors.

## 5 Funding

This research was funded by the National Science Fund for Distinguished Young Scholars of China under grant number: 62325308; the Science and Technology Innovation 2030–New Generation Artificial Intelligence Major Project under grant number No. 2018AAA0100103; the National Natural Science Foundation of China under grant number 62002297, grant number 61722212, grant number 62072378, grant number 62273284, grant number 62472353, grant number 62302495, and grant number 62172338; the Neural Science Foundation of Shaanxi Province under grant number: 2022JQ-700; the Fundamental Research Funds for the Central Universities under grant number: D5000230199; the Guangdong Basic and Applied Basic Research Foundation under grant number 2024A1515011984; and the Fundamental Research

Funds for the Central Universities under grant number G2023KY05102.

## 6 Declaration of competing interest

The authors declare that they have no conflicts of interest in this work.

**Supplementary File 1**

**Updating *θ*.** To update $\theta$, we fix $S$ and eliminate terms that do not pertain to $\theta$. Consequently, the objective function in Equation (5) simplifies to $\mathcal{L}_{Cls}(\theta, A, X, Y)$, which we can simply learn $\theta$ via stochastic gradient descent.

**Updating *S*.** To update $S$, we fix $\theta$ and eliminate terms that do not pertain to $S$. Consequently, the objective function in Equation (5) simplifies to:

$$\min_{S} \mathcal{L}(S, A) + \alpha \|S\|_1 + \beta \|S\|_* \quad s.t., \ S = S^T \tag{1}$$

where $\mathcal{L}(S, A) = \|A - S\|_F^2 + \gamma \mathcal{L}_{Cls} + \lambda \mathcal{L}_{graph}$ (2)

It should be noted that both the ℓ1 norm and the nuclear norm are inherently non-differentiable. In scenarios where the optimization problem incorporates a singular non-differentiable regularizer $R(S)$, the application of Forward-Backward splitting methods [1] is advisable. This approach alternates between a gradient descent step and a proximal step, as described below:

$$S^{(k)} = \text{prox}_{\eta R}\left(S^{(k-1)} - \eta \nabla_S \mathcal{L}(S, A)\right), \tag{2}$$

where $\eta$ represents the learning rate, and prox$R$ denotes the proximal operator, defined as follow:

$$\text{prox}_R(Z) = \arg\min \frac{1}{2} \|S - Z\|_F^2 + R(S) \tag{3}$$

Specifically, the proximal operators for the ℓ1 norm and the nuclear norm can be expressed as:

$$\text{prox}_{\alpha\|\cdot\|_1}(Z) = sgn(Z) \odot (|Z| - \alpha)_+ \ , \tag{4}$$

$$\text{prox}_{\beta\|\cdot\|_*}(Z) = \text{U}diag((\sigma_i - \beta)_+)_i V^T \ , \tag{5}$$

where $Z = \text{U}diag(\sigma_1, \ldots, \sigma_n) V^T$ is the singular value decomposition of $Z$. To manage the optimization of an objective function that includes two non-differentiable regularizers, Richard et al. [2] propose the Incremental Proximal Descent method, utilizing the aforementioned proximal operators. This method involves cyclically iterating the update process, allowing for the updating of $S$ as described below:

$$\begin{cases} S^{(k)} = S^{(k-1)} - \eta \cdot \nabla_S\big(\mathcal{L}(S, A)\big), \\ S^{(k)} = \text{prox}_{\eta\beta\|\cdot\|_*}\big(S^{(k)}\big), \\ S^{(k)} = \text{prox}_{\eta\alpha\|\cdot\|_1}\big(S^{(k)}\big), \end{cases} \tag{6}$$

Once we have learned a relaxed version of $S$, we project it to satisfy specific constraints. For the symmetry constraint, we define $S$ as $S=(S+S^T)/2$. Regarding the constraint that each element $S_{ij}$ must lie within the range [0, 1], we project any $S_{ij}$ values less than 0 to 0 and any greater than 1 to 1. We denote these projection procedures as $P_S(S)$.

**Training Algorithm:**

***Initialize*** $S \leftarrow A$

***Randomize*** $\theta$

***while*** the error has not converged ***do***

$S \leftarrow S - \eta \cdot \nabla_S \left( \|A - S\|_F^2 + \gamma \mathcal{L}_{Cls} + \lambda \mathcal{L}_{graph} \right)$

$S \leftarrow \mathrm{prox}_{\eta\beta\|\cdot\|_*}(S)$

$S \leftarrow \mathrm{prox}_{\eta\alpha\|\cdot\|_1}(S)$

$S \leftarrow P_S(S)$

***for*** $i$ = 1 to τ ***do***

$g \leftarrow \frac{\partial \mathcal{L}_{Cls}(\theta, A, X, Y)}{\partial \theta}$

$\theta \leftarrow \theta - \eta' g$

***Return*** $S$, $\theta$